\documentclass[a4paper]{spie}  %>>> use this instead for A4 paper
\usepackage{amsmath,amsfonts,amssymb}
\usepackage{graphicx}
\usepackage{hyperref}
\hypersetup{colorlinks=true, allcolors=blue}
\usepackage[dvipsnames,svgnames,x11names,table]{xcolor}
\usepackage{comment}

\newcommand{\highrisk}[1]{{\bf \color{BrickRed} X}}
\newcommand{\modrisk}[1]{{\bf \color{YellowOrange} -}}
\newcommand{\lowrisk}[1]{{\bf \color{ForestGreen} $\bm{\checkmark}$}}

\title{The Atacama Large Aperture Submillimetre Telescope (AtLAST)}

\author[a]{Pamela D. Klaassen}
\author[b]{Tony Mroczkowski}
\author[c]{Claudia Cicone} 
\author[b]{Evanthia Hatziminaoglou}
\author[d]{Sabrina Sartori}
\author[b]{Carlos De Breuck}
\author[e]{Sean Bryan}
\author[f]{Simon R. Dicker}
\author[b,g]{Carlos Duran}
\author[d]{Chris Groppi}
\author[h]{Hans K\"archer}
\author[ijk]{Ryohei Kawabe}
\author[lm]{Kotaro Kohno}
\author[n]{James Geach}

\affil[a]{UK Astronomy Technology Centre, Royal Observatory Edinburgh, Blackford Hill, Edinburgh EH9 3HJ, UK}
\affil[b]{European Southern Observatory (ESO), Karl-Schwarzschild-Strasse 2, Garching 85748, Germany}
\affil[c]{Institute of Theoretical Astrophysics, University of Oslo, P.O. Box 1029, Blindern, 0315 Oslo, Norway}
\affil[d]{Department of Technology Systems, University of Oslo, NO-2027 Kjeller, Norway}
\affil[e]{School of Earth and Space Exploration, Arizona State University, Tempe, AZ 85287 USA}
\affil[f]{Department of Physics and Astronomy, University of Pennsylvania, 209 South 33rd Street, Philadelphia, PA, 19104, USA}
\affil[g]{European Southern Observatory (ESO), Santiago, Chile}
\affil[h]{Independent Contractor,  Kirchgasse 4, D-61184 Karben, Germany}
\affil[i]{Department of Astronomy, The University of Tokyo, 7-3-1, Hongo, Bunkyo-ku, Tokyo, 113-0033, Japan}
\affil[j]{National Astronomical Observatory of Japan Mitaka, Tokyo 181-8588, Japan}
\affil[k]{The Graduate University for Advanced Studies (SOKENDAI), 2-21-1 Osawa, Mitaka, Tokyo 181-8588, Japan}
\affil[l]{Institute of Astronomy, The University of Tokyo, 2-21-1 Osawa, Mitaka, Tokyo 181-0015, Japan}
\affil[m]{Research Center for the Early Universe, School of Science, The University of Tokyo, 7-3-1 Hongo, Bunkyo, Tokyo 113-0033, Japan}
\affil[n]{Centre for Astrophysics Research, School of Physics, Astronomy \& Mathematics, University of Hertfordshire, Hatfield, AL10 9AB, UK}

\authorinfo{Further author information: (Send correspondence to P.D.K.)\\P.D.K.: E-mail: pamela.klaassen@stfc.ac.uk, Telephone: +44 (0)131 668 8218\\ For further information or to sign up for the working groups, please see \url{https://atlast-telescope.org}}

\begin{document} 
\maketitle

\begin{abstract}
The coldest and densest structures of gas and dust in the Universe have unique spectral signatures across the (sub-)millimetre bands ($\nu \approx 30-950$~GHz).
The current generation of single dish facilities has given a glimpse of the potential for discovery, while sub-mm interferometers have presented a high resolution view into the finer details of known targets or in small-area deep fields. However, significant advances in our understanding of such cold and dense structures are now hampered by the limited sensitivity and angular resolution of our sub-mm view of the Universe at larger scales.

In this context, we present the case for a new transformational astronomical facility in the 2030s, the Atacama Large Aperture Submillimetre Telescope (AtLAST). AtLAST is a concept for a 50-m-class single dish telescope, with a high throughput provided by a 2~deg - diameter Field of View, located on a high, dry site in the Atacama with good atmospheric transmission up to $\nu\sim 1$~THz, and fully powered by renewable energy.

We envision AtLAST as a facility operated by an international partnership with a suite of instruments to deliver the transformative science that cannot be achieved with current or in-construction observatories. As an 50m-diameter telescope with a full complement of advanced instrumentation, including highly multiplexed high-resolution spectrometers, continuum cameras and integral field units, AtLAST will have mapping speeds hundreds of times greater than current or planned large aperture ($>$ 12m) facilities. By reaching confusion limits below L$_*$ in the distant Universe, resolving low-mass protostellar cores at the distance of the Galactic Centre, and directly mapping both the cold and the hot (the Sunyaev-Zeldovich effect) circumgalactic medium of galaxies, AtLAST will enable a fundamentally new understanding of the sub-mm Universe.
\end{abstract}

% Include a list of keywords after the abstract (up to 10)

%\section*{Publications to cite - please add!!!}
% put links here to papers I should cite in the proceedings.

%the Large Latin American Millimeter Array (LLAMA)\cite{Romero2020}

\section{INTRODUCTION}
\label{sec:intro}  % \label{} allows reference to this section

ALMA \cite{Wootten2009} and other recently upgraded interferometers, such as NOEMA and the SMA, have revolutionised our understanding of the sub-mm sky. Across many astronomical disciplines, from detection and imaging of the atomic and molecular interstellar medium at the highest redshifts \cite{Hashimoto+18} %[OIII] 88micron at z=9.1
\cite{Rizzo2020} 
and resolving ring structures in protostellar disks\cite{HLTau2015}, to providing collecting area
for very-long-baseline interferometry (VLBI) campaigns that delivered the first images of a black hole \cite{EHT2019}, we have gained a much deeper understanding of how the hidden Universe evolves.

With observations at sub-mm wavelengths, we are able to probe many phases of baryonic matter, from the cold dense medium directly detected in emission or absorption, to the hot ionised gas revealed through the Sunyaev-Zeldovich effect. Interferometers are crucial for understanding the detailed physics involved.  However, by their very nature they, on the one hand  filter out the emission on the large scales, and on the other, are not suited to perform the deep, wide field surveys that can provide statistically significant samples of targets because of their small fields of view. Indeed, without a deep, high resolution map of the sub-mm sky,  eventually ALMA will become \textit{source starved:} any detailed follow-up studies will remain limited to the targets whose positions were determined through past -  sensitivity-limited, low-resolution - surveys, probing only the `tip of the iceberg' of any population of sources relevant for star formation and galaxy evolution studies. This means there is a crucial gap in survey capabilities at these wavelengths, and in the era of ALMA, SKA and ngVLA, it becomes more and more important to fill.

Closing this gap requires a new, large aperture single-dish observatory, with a wide field of view (FoV) and a sufficiently accurate surface to enable (sub-)mm observations with good aperture efficiency.
A field of view larger than any large aperture ($>10$-m) telescope currently under construction, combined with a large collecting area required for sensitivity and spatial resolution, will deliver a high throughput telescope capable of hosting highly multiplexed instrumentation. The need for a large aperture is driven by scientific requirements such as overcoming the confusion limit in high-z surveys, and resolving nearby protostellar cores, as well as achieving sufficient overlap with ALMA/ACA in visibility ($uv$) space. 
%% In particular, this new telescope must have a resolution comparable to that of the Atacama Compact Array (ACA)\cite{Iguchi2009}, the 7-metre, 12-element compact component of ALMA with a maximum baseline of $\sim$ 45m. \tony{I would say that it's a nice feature, and that a convenient way of thinking of AtLAST is like the ACA but fully filled and with a FoV that will be thousands of times larger than the ACA in Band 3... but that it's not driving the requirement.  the requirement is more the resolution for high-z and Galactic studies, and achieving sufficient overlap with ALMA and the ACA in uv-space (mostly the former, due to the poor mapping speed of the ACA).} %%% TM: I added something to this effect in the conclusions 
%\cc{I agree that statement was confusing, I have slightly rephrased the previous one and commented out the original statement}
Such a facility has been identified as a priority in the long range plans\footnote{http://www.eso.org/sci/facilities/alma/announcements/20180712-alma-development-roadmap.pdf} of a broad scientific community that includes many users of ALMA and other sub-mm facilities. 
%but is out of scope for the ALMA agreements \cc{Do we need to specify this last sentence?}.
Further, this new infrastructure needs to be placed on a high, dry site with optimal atmospheric transmission at sub-mm wavelengths, but at the same time it must be safe to access and operate. Given the global crises we now face, and in order to meet the societal and climate challenges of the future, this new facility must mitigate its impact on the natural environment of the pristine Atacama desert, and it must have a positive and inclusive societal effect on the local and international communities that it serves.

Here we outline the overall case for such a facility: the Atacama Large Aperture Submillimetre Telescope (AtLAST)\cite{Klaassen2019}.
We also discuss the developing key science cases for the facility and a recently-approved EU-funded project\footnote{\url{https://cordis.europa.eu/project/id/951815}} that will generate a workable and comprehensive design study and budget for the AtLAST observatory. 
%This study, funded through the EU Horizon 2020 Research and Innovation ``Design Studies'' programme, aims to create a comprehensive and costed telescope design for AtLAST.
%a 50-m-class single dish observatory with a large field of view and multiple dedicated facility instruments able to do survey, target of opportunity (TOO) and PI science.

%https://cordis.europa.eu/project/id/951815

\section{SCIENCE DRIVERS}
\label{sec:science}

There is a sizeable gap in the parameter space that astronomers are able to probe in both sensitivities and scales with current sub-mm telescopes: interferometers can see small-scale structures in exquisite detail, and current single dish facilities can pick up the large-scale structures by {\it sampling} wider fields of the sky. What is not feasible with current facilities is to generate statistical understandings of the source populations at the spatial resolution required to resolve smaller details and so de-blend compact structures.
%and statistics from which those single dish surveys are pulled, 
%the ability to combine the large-scale information with the detailed high resolution view.  

The throughput we propose for AtLAST will enable large-scale surveys at multiple wavelengths that will give definitive catalogues of the cold Universe.  With a 50m diameter and 2$^\circ$ FoV, we could undertake a sub-mm SDSS-style extragalactic survey down to L$_*$ sensitivities before hitting the confusion limit\cite{Geach2019}. We could similarly map the interstellar medium in the Galactic Plane, capturing dust from protostellar cores on 0.1 pc size scales at the distance of the Galactic Centre, while using CH$_3$CN as a temperature probe across the bulk of the Galaxy\cite{Stanke2019}. To properly carry out such surveys, assuming at least three passes at different wavelengths to build up spectral energy distributions (SEDs) and spectral line energy distributions (SLEDs) would only take a few years (e.g. $<$ 10 yr) with a suitably instrumented AtLAST. To reach the same sensitivities and sky coverage would take ALMA millennia to complete, while still not recovering the large-scale structures. 

To serve the whole community however, the AtLAST observatory must be much more than just a survey instrument and {\it it must perform PI science}. For instance, time variable star formation in nearby regions cannot be monitored efficiently using current facilities \cite{Fischer2019} on the size scales of the variability, nor can the large scale structures of the circumgalactic medium (CGM) of nearby galaxies - a key probe of galaxy evolution - be properly measured\cite{Cicone2019}.

%% protostellar variability (written by Pamela)

ALMA's key science goal of resolving protostellar disks has opened untold richness in understanding planet formation, but there is still a lot to learn about how the central protostar(s) in these disks accrete their material. Quantifying protostellar variability can help uncover the physical processes involved in accreting material onto these stars. Accretion is thought to be an episodic process, with accretion bursts accounting for most of the material deposited on stars.  As material accretes onto the forming star, it heats up, and these accretion bursts can be detected through flaring at various wavelengths.  The shorter wavelength emission flares trace the inner disk regions, and how those flares propagate to other wavelengths, both in time and in in amplitude gives significant insights on the structures within these disks\cite{Fischer2019}.  Studies like these require a large, high-throughput telescope for the resolution and multiplexing they allow: enabling fast mapping to capture large portions of a star forming region in multiple epochs relatively quickly.

%% Galactic plane science (written by Pamela)

While sensitivity is not a key factor in the above studies of nearby star forming regions, it becomes a factor for determining the locations of star forming regions across the Galactic Plane.  In the coming decades, sensitive maps of the dust and line emission from the dense interstellar medium (ISM) will be required to study the statistics of star formation in our Galaxy.  Mid-J CO and atomic [CI] line surveys, along with continuum surveys, will allow us to probe star formation across the entire Galaxy and understand the roles of filaments and fibres in the creation of star forming regions\cite{Stanke2019}. With atomic and molecular line surveys of the plane, we can also probe photon dominated regions (PDRs) at the interface of star clusters and their natal environments\cite{Goicoechea2016}, and even probe the effects on the next generation of stars\cite{Klaassen2020} in a comprehensive way. And through studies like this, we understand more about the evolution of our Galaxy as a whole.

%%% CGM Science Case (written by Claudia)

%It has now become clear that galaxies at all redshifts are embedded in massive gaseous haloes extending by up to $\sim100$~kpc\cite{Wisotzki+18}. 
The CGM properties are directly shaped by the cycling of baryons in and out of galaxies\cite{Suresh+15,Sorini+20,Stern+20}, and so provide a unique tool to test the assumptions of current galaxy formation and evolution theories. Although most models predict the CGM to be dominated by hot rarefied gas with $T\geq10^5$~K, in high redshift galaxies observed using deep and well sampled interferometric observations, the CGM was - quite surprisingly - found to include a significant fraction of molecular gas extending up to tens of kpcs\cite{Cicone+15,Emonts+16,Ginolfi+17,Fujimoto+20}. The formation of such massive cold CGM reservoirs may be aided by massive molecular outflows\cite{Cicone+18, DiTeodoro+20} and/or galaxy interactions within a dense environment\cite{DeBlok+18}. However, because of their low surface brightness and wide angular extent - molecular CGM components would escape observations conducted with current facilities. Especially in the local Universe, this emission is on scales too large for ALMA, and most of it is too faint to be detected by current single-dish telescopes.

%% galaxy surveys (written by Jim)

A large area (1000\,deg$2$) low confusion `cosmological' survey conducted with AtLAST that combines photometric and low-resolution ($R$ equivalent to $\sim$100\,km\,s$^{-1}$) spectroscopic components would enable transformative extragalactic science; we envision sub-mm analogue to SDSS\cite{Geach2019}. By measuring the redshifts, star formation rates and dust masses of hundreds of thousands of galaxies down to luminosities equivalent to $L_\star$ across the bulk of cosmic time, it will be possible to make a 3D map of galaxy evolution right out to $z\approx10$, well into the epoch of reionisation. This is vital for our understanding of galaxy evolution since half of the radiation emitted by stars over cosmic history has been absorbed and re-emitted by dust. The unrivalled sensitivity and resolution of AtLAST will: provide a complete census of star-forming galaxies during the `ramp up' phase of galaxy growth from cosmic dawn ($z\sim10$) to cosmic noon ($z\sim1$--$2$); reveal the production and evolution of metals as traced by dust; directly probe the evolution of the co-moving molecular gas density in galaxies from early times, thus testing models of gas accretion and star formation efficiency. Beyond these galaxy evolution studies, such a survey will map the growth of large scale structure within which the galaxies are forming and evolving, providing unique constraints on cosmological parameters, for example through the measurement of baryonic acoustic oscillations beyond $z>2$. Thus, AtLAST will provide complementarity to large area galaxy surveys such as Euclid and LSST.

%%% SZ science (written by Tony)
While many of the main goals of AtLAST relate to cold molecular gas, (sub)mm-wave observations also offer the ability to probe the warm ($10^5-10^7~\rm K$) and hot ($>10^7~\rm K$) ionised gas pervading the intergalactic and circumgalactic medium.
%large scale structure throughout the Universe.  
Such gas is typically invisible to optical telescopes, and traditionally has required space-borne X-ray observatories, but the thermal and kinetic Sunyaev-Zeldovich (SZ) effects \cite{Sunyaev1972, Sunyaev1980} provide alternative, complementary views.
Both the thermal and kinetic SZ (tSZ and kSZ) effect are redshift independent distortions of the CMB, with unique continuum spectra that are separable when probed from frequencies $\sim30-500$~GHz.
The dominant thermal SZ effect provides a measure of the thermal energy content of the gas within a beam.  In low resolution surveys, this is useful for identifying large scale structures like galaxy clusters, where -- because the gas is close to virial equilibrium and supported by thermal pressure --  the thermal energy is a good proxy for total mass \cite{Carlstrom2002,Motl2005}.  At higher resolution, the thermal SZ provides a probe of cluster astrophysics \cite{Mroczkowski2019, Mroczkowski2019b}.  The kinetic SZ effect, generally subdominant to the thermal for high temperature ($\gtrsim 1~\rm keV$), is a Doppler shift of the CMB that probes the product of motion along the line of sight and electron opacity.  Recent studies have shown the kSZ to be a powerful probe of ionised CGM gas down to galaxy scales \cite{Amodeo2020, Chiang2020, Schaan2020}.
It is worth noting that both the thermal and kinetic SZ effects require relativistic corrections when the gas is extremely hot or in a non-thermal or non-Maxwellian distribution \cite{Itoh1998, Itoh2004, Pfrommer2005, Chluba2014a, Chluba2014b, Basu2019, Mroczkowski2019}; these corrections each uniquely affect the overall SZ spectrum, and offer the unique opportunity to provide a ground-based view of cluster temperature and exotic plasma composition that normally may only be probed at X-ray and $\gamma$-ray energies from space.
High spatial and spectral resolution observations of the SZ effects are therefore strongly among the key science drivers for AtLAST.

Combining the above and several other transformational science cases motivate the need for a new, larger single-dish facility like AtLAST, and as part of the Horizon 2020 Design Study, we will be refining them into science requirements for the facility.

\section{CURRENT STATE OF THE ART}
\label{sec:current_facilities}

ALMA\cite{Wootten2009} is currently the largest ground-based sub-mm astronomical observatory in the world, and has had enormous success achieving its original `level-one' science goals,\cite{DeBreuck2005} which were: detecting CO or C$^+$ in a Milky Way like galaxy at a redshift of $z=3$, imaging protoplanetary disks in solar like systems at a distance of 150 pc\cite{HLTau2015,Itziar2013,Qi2013}, and producing stunning images of astronomical phenomena at resolutions finer than 0.1$''$. ALMA now routinely accomplishes all these science goals.

In the Northern Hemisphere, \hyperlink{https://www.iram-institute.org/EN/noema-project.php?ContentID=9&rub=9&srub=0&ssrub=0&sssrub=0}{NOEMA}  is also having tremendous success as it builds up to its full complement of 12 antennas and increases its baseline lengths\cite{Fontani2017,Beuther2018}, and the \hyperlink{https://www.cfa.harvard.edu/sma/}{SMA} is fully taking advantage of its unparalleled correlator bandwidths to deepen our understanding of the sub-mm universe.

Nevertheless, by the very nature of interferometry, these observatories are unable to capture the emission from large scale structure, which requires single-dish facilities.  Nor can inteferometers carry out large area unbiased surveys nearly as efficiently as their single-dish counterparts.  
In this area, facilities such as the \hyperlink{https://www.eaobservatory.org/jcmt/}{JCMT} and the \hyperlink{https://www.iram-institute.org/EN/30-meter-telescope.php}{IRAM 30m} have been sharpening our understanding of the sky for decades, with newer facilities such as \hyperlink{http://www.apex-telescope.org/ns/}{APEX} and the \hyperlink{http://lmtgtm.org}{LMT} adding to our understanding of large scale structures in the sub-mm sky thanks to their high/dry sites and sensitive instrumentation suites. Many of these facilities have receivers making them capable of joining Very Long Baseline interferometers like the \hyperlink{https://science.nrao.edu/facilities/vlba}{VLBA} and \hyperlink{www.evlbi.org}{EVN} networks and the \hyperlink{https://eventhorizontelescope.org}{EHT}. Indeed, this is expected to be a primary working mode for LLAMA\cite{Romero2020}. 

Similarly, smaller aperture experiments like {\it Planck}, ACT, and SPT, which were built primarily for wide area surveys of the CMB, have delivered tremendous -- and often unforeseen -- possibilities through their large, unbiased sky surveys and rich datasets.
On the horizon are a number of similarly-sized facilities for the highest reaches of Cerro Chajnantor such as \href{http://www.ioa.s.u-tokyo.ac.jp/TAO/en/}{TAO} and \href{https://www.ccatobservatory.org}{FYST}.
We also note the Africa Millimetre Telescope (\href{https://www.ru.nl/blackhole/africa-millimetre-telescope/}{AMT}) in Namibia will also deliver single dish observations at a resolution similar to that of the JCMT, though likely will primarily operate in lower mm-wave bands due to its lower elevation (2350~m above sea level).

\section{The need for a new facility}
\label{sec:why_atlast}

The next big leap in the fields of sub-mm astronomy will require large-aperture, wide field of view single dish telescopes such as AtLAST.  There are two key areas where current large-aperture facilities fall short: they cannot deliver highly-populated (highly multiplexed) large focal planes, and they do not reach the crucial size scales just resolved out by ALMA and the other interferometers \cite{Frayer2017}. Combining these capabilities will give us an unprecedented and comprehensive panoramic view of the sub-mm sky from the Solar System to large scale structures across the Universe.

The longest baselines of the ALMA Compact Array (ACA, or Morita Array) are roughly 45~m, and all of the single dish facilities mentioned above (with the exception of the LMT) are much smaller than that, and so their resolution is proportionately coarser, as are their relative sensitivities.  With a single 50-m-class dish, we will deliver a collecting area $>4\times$ that of the full 12-element ACA (and thus $\gtrsim4\times$ better point source sensitivity) while also delivering comparable resolution. 
In addition to this sensitivity improvement, AtLAST will have the advantage of a much greater FoV ($\sim$ 2$^\circ$ as opposed to $\sim$30$^{\prime\prime}$), which, with a fully populated focal plane, makes it much more efficient than the ACA or even full ALMA in terms of mapping speed, while at the same time not suffering from filtering of the large scale structures of interest.

Planned upgrades to ALMA will see its capabilities (such as imaging speed or resolution) improved by a factor of a few, but will not improve its field of view, or sensitivity to large scale structures. Only a new facility can deliver on those requirements. It is only with a larger FoV, wide aperture telescope that we can hope to meet the science challenges posed above in Section \ref{sec:science}.

Many ground and space based missions planned into the 2030s are motivated by complementary scientific goals to those of AtLAST, including the \hyperlink{https://www.eso.org/public/teles-instr/elt/}{ELT}, \hyperlink{https://www.skatelescope.org}{SKA}, \hyperlink{www.jwst.nasa.gov}{JWST} and \hyperlink{https://www.the-athena-x-ray-observatory.eu}{ATHENA}.  These facilities share many common interests, but address them from unique perspectives.  Additionally, large dedicated survey telescopes are also expected to come online in the next decade, including the Rubin Observatory, Euclid and the Nancy Grace Roman Telescope. They approach many of the same scientific questions from different wavelengths, and therefore the contributions of AtLAST to these studies will be unique: AtLAST will be the only observatory that can provide large scale survey data in the sub-mm bands.

\begin{figure}
    \centering
    \includegraphics[width=0.99\textwidth]{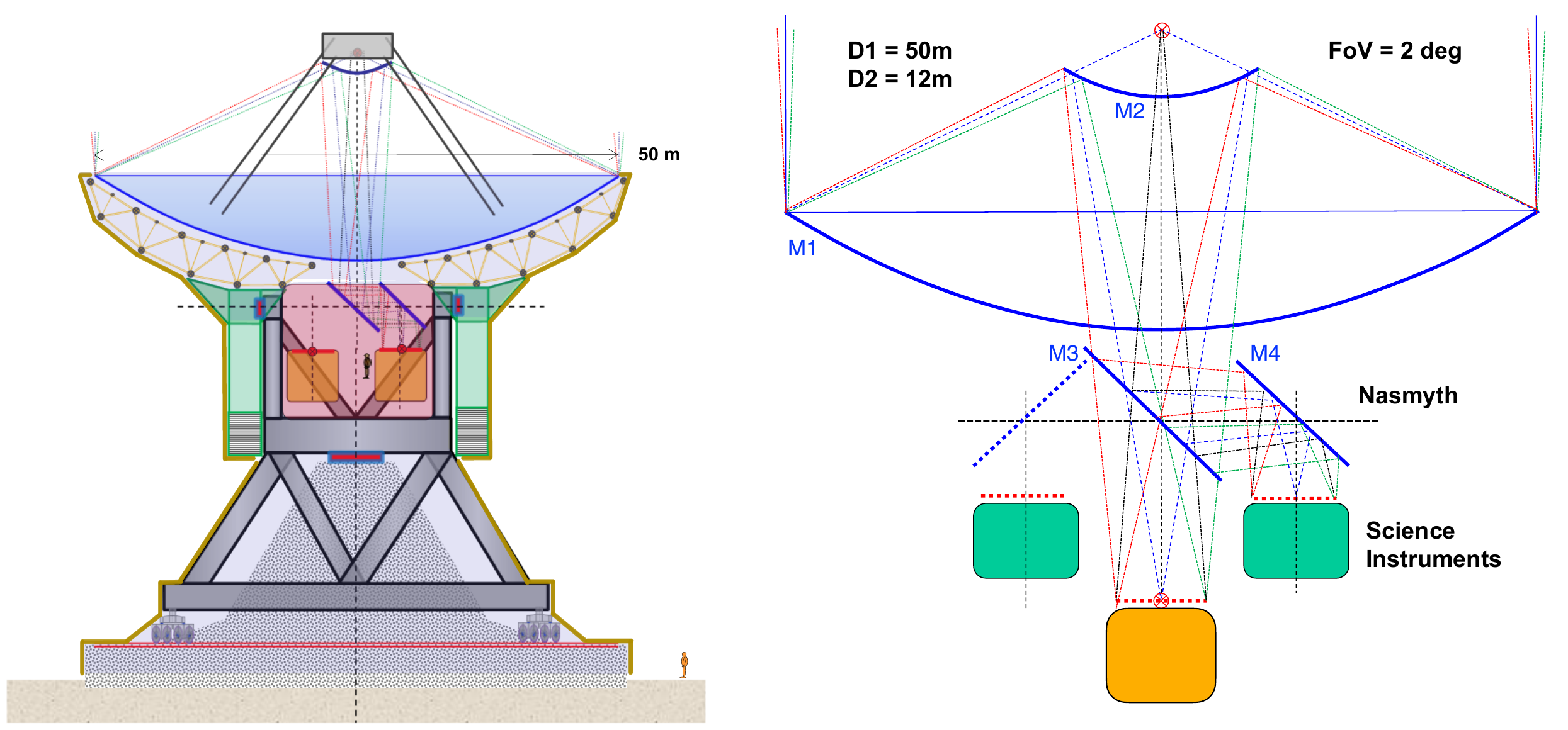}
    \caption{Baseline telescope design for AtLAST.  The left panel depicts a concept for the mechanical structure of the on-axis Cassegrain-Nasmyth hybrid design.  Adult humans are depicted for scale in the receiver cabin and to the right of the telescope base.
    The right panel depicts the baseline optical concept.  The primary (M1), secondary (M2), and tertiary (M3) move with the elevation axis and are generally held fixed relative to each other when observing with one of 4-6 Nasmyth-mounted receivers/instruments (green boxes).  The quaternary mirror, M4, moves only in azimuth with the entire telescope structure (i.e.\ along with the instruments and M1-3).  
    To observe with the Cassegrain-mounted instrument (depicted as an orange box), M3 is moved out of the optical path (i.e.\ to the dashed blue line).  The Cassegrain-mounted instrument moves in elevation, fixed relative to M1 and M2. Note the scales differ in the two panels, though the 50-m primary mirror aperture serves as a scale reference for each.}
    \label{fig:atlast_design}
\end{figure}

\section{THE ATLAST PROJECT}
\label{sec:atlast}

%\pk{Kotaro: Is this maybe a good place to mention the Japanese Metrology studies, and to reference your SPIE paper?}
% TM: added citation to LST below.  for the metrology, we could elaboration quite a bit and include Hans' paper, LASSI, and Nobeyama like we did in the proposal

Our baseline design is for a 50-m on-axis Cassegrain-Nasmyth hybrid configuration (see Figure \ref{fig:atlast_design}).  While several off-axis designs for smaller (1--10 m) telescopes exist and provide wide fields and unblocked apertures, we feel the challenge of delivering a sufficiently rigid structure that can achieve fast enough scan speeds to modulate the atmosphere sufficiently rapidly ($\gtrsim 0.2~\rm Hz$, or $\gtrsim 1 \rm deg~s^{-1}$) is best met by building on the experience of the 50-m LMT design: There are significant  similarities to the 50-m LMT \cite{Hughes2010}, and indeed the 50-meter Large Submillimeter Telescope (LST) concept \cite{Kawabe2016}, in the left panel of Figure \ref{fig:atlast_design}.  
Aside from a higher and drier site and access to more of the Southern sky, some key advantages to AtLAST are a large, 12-m secondary mirror enabling $\sim 500\times$ the FoV of the LMT, and an improved receiver cabin to host larger and more instruments from the start.  

\begin{table}[htbp]
\caption{Telescope Characteristics Table}
\label{tab:telescope}
\centering
\begin{tabular}{|l|l|l|}
\hline
%\multicolumn{2}{c!{\color{tabclr}\vrule}}{TBC }       \\
Telescope Property	    & Expected Value	&Units\\
\hline
Main and Effective Aperture Size		& 50      & m       \\
System Effective Focal Length		    & $\sim 100$               &   m    \\
Sizes of Segments		                & $\approx 10$              & m$^2$      \\
Number of Segments	                    & $\approx 200$          &       \\
Total Collecting Area		            & $\approx$2000           & m$^2$      \\
Field of View		                    & goal=2, min=1        &  deg     \\
Wavelength range		                & 0.3-10 & mm      \\
Optical surface figure quality (RMS)	& 20-25             & $\mu$m       \\
Surface Coating Technique		        & similar to ALMA             &       \\
Number of Mirrors                       & 2-4              &       \\		
Actuator Precision 		    & $\approx 10$               &  $\mu$m     \\
Mount Type                              & altitude-azimuth (azimuth-elevation) & \\		
Support Structure Material              & steel \& invar              &       \\ 		
Optical Design       & hybrid Cassegrain-Nasmyth            &       \\
Number of Instruments                   & 1 Cassegrain, 4 Nasmyth            &       \\		
Description of Adaptive Optics		    &  TBD, active surface  & \\ 
\hline
\end{tabular}
\end{table}

\begin{figure}
   \centering
   \includegraphics[angle=270,width=0.95\textwidth,clip,trim=20mm 4mm 60mm 124mm]{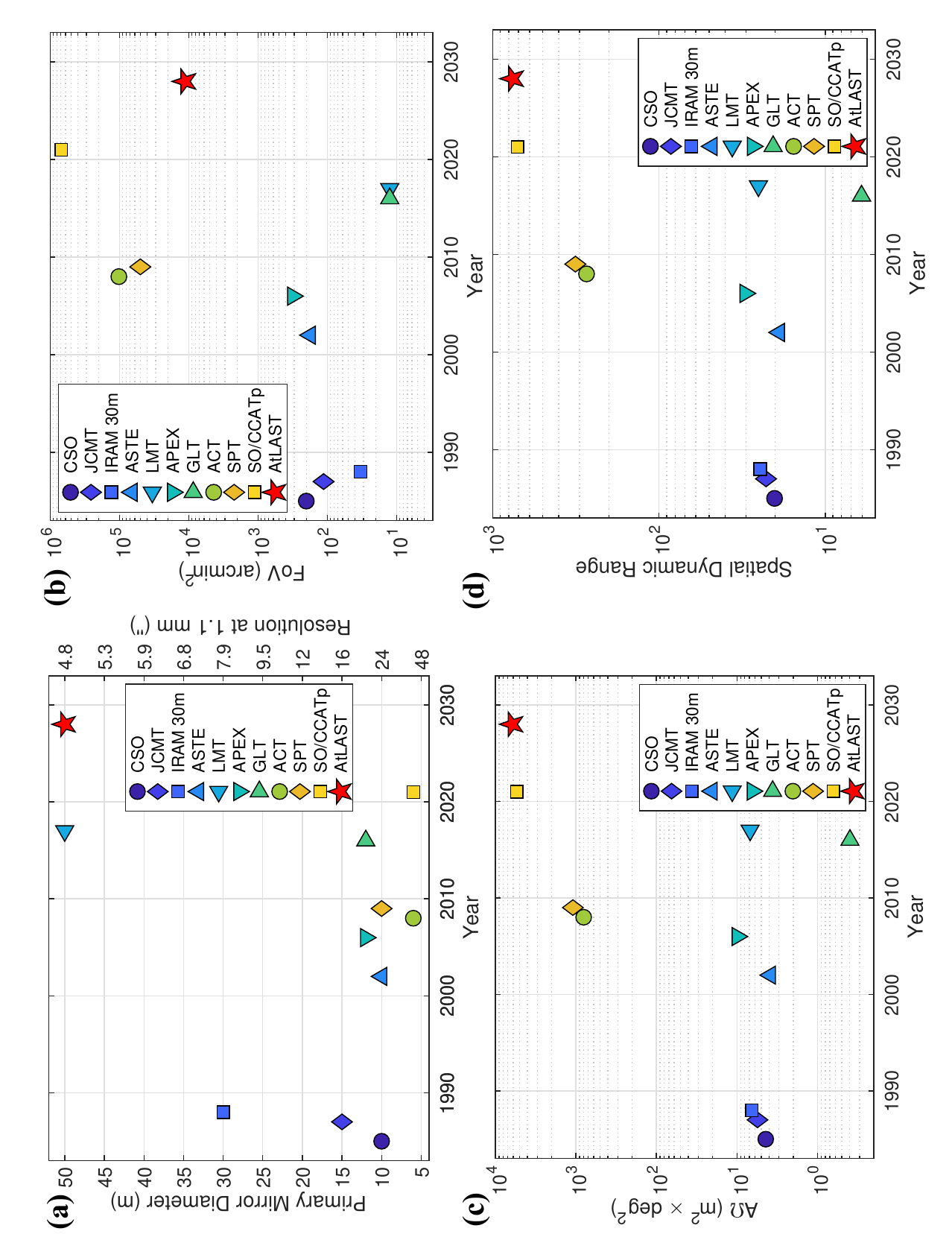}
   \caption{Primary dish diameter and spatial resolution (panel a), field of view (panel b), throughput (A$\Omega$ = FoV $\times$ collecting area, also known as étendue, panel c), and  spatial dynamic range (defined as the number of resolution elements across the FoV, panel d) of existing or funded single-dish telescopes operating at (sub)mm wavelengths. Only facilities that can (or will) observe frequencies of up to at least 270 GHz (1.1mm) are included. Acronyms, defined for convenience in Table \ref{tab:acronyms} in the appendix, correspond to: Caltech Submm Observatory (CSO); James Clerk Maxwell Telescope (JCMT); the 30-m telescope of the Institut de radioastronomie millimétrique (IRAM); the Atacama Submillimeter Telescope Experiment (ASTE);  The Large Millimeter Telescope or Gran Telescopio Milimétrico (LMT/GTM);  the Atacama Pathfinder Experiment (APEX);  the Greenland Telescope (GLT), the Atacama Cosmology Telescope (ACT); the South Pole Telescope (SPT); the Simons Observatory (SO) and CCAT-prime (which share the same design), and our concept for the Atacama Large Aperture Submillimetre Telescope (AtLAST, which is marked with a red star).}
   \label{fig:telescope}
\end{figure} 

Like most large single dish telescopes, our baseline telescope design for AtLAST will include an active surface.  From the start, we will included closed-loop metrology that will largely bypass the need for routine astronomical holographic focusing procedures. These procedures  can often incur a significant observational overhead and reduce the beam quality and forward gain by an often unknown and systematic amount.  The precise implementation is for the metrology system is one of the topics being considered in the Horizon 2020 study discussed below. We note that a number of viable solutions exist, such as millimetric adaptive optics (MAO; PI Tamura) being tested on the Nobeyama 45-m Radio Telescope\cite{Tamura2020}, flexible body control (FBC)\cite{Kaercher2006} and the Laser Active Surface Scanning Instrument (LASSI)\cite{SeymourLockman2019,Salas2020} currently being implemented on the 100-m Green Bank Telescope (GBT).

There are a number of baseline assumptions we have made with respect to the AtLAST project, not the least of which are that we envision a standalone observatory capable of both large surveys and user-proposed observations, that the telescope requirements (given in Table \ref{tab:telescope}) are driven by key science goals, and that there will be a focus on conscientious and sustainable technology development. 
Figure \ref{fig:telescope} compares this baseline concept for AtLAST to several other major mm-wave observatories, highlighting its design goal to provide superior throughput and dynamic range (panels c and d of Figure \ref{fig:telescope}).  Note that for simplicity, we restrict our comparison to telescopes capable of observations at 1.1~mm.

\begin{figure}
    \centering
    \includegraphics[width=0.95\textwidth]{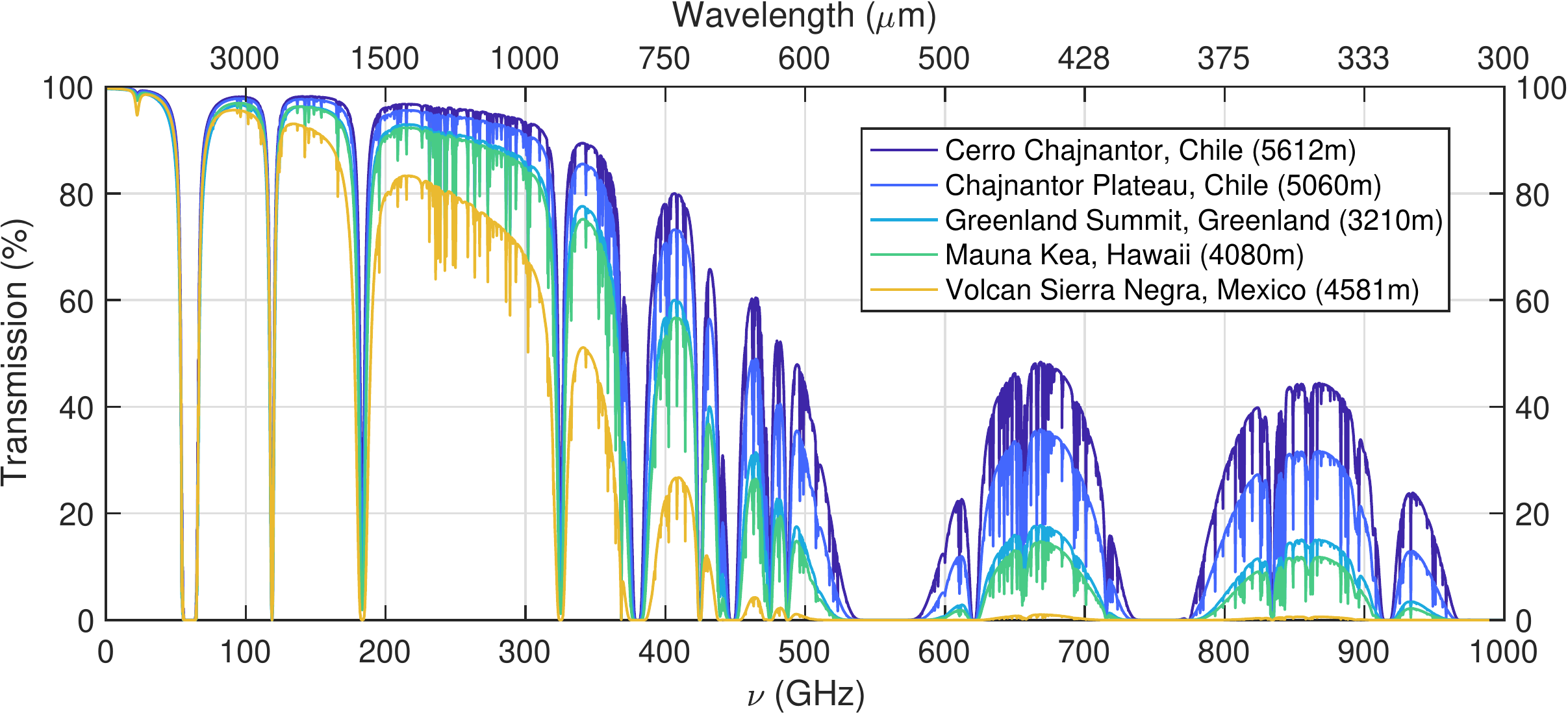}
    \caption{Median transmission at several premiere sites used in millimeter/submillimeter astronomy.
    Cerro Chajnantor is the location of the CCAT-prime observatory.
    The Chajnantor Plateau hosts observatories such as ALMA and APEX, and offers conditions nearly comparable to those near the summit of Cerro Chajnantor.
    The Greenland Summit is the future site of the GLT.
    Mauna Kea is host to the SMA and JCMT, along with many optical facilities.
    Volcan Sierra Negra is the site of the GTM/LMT in Puebla, Mexíco.
    AtLAST will likely be located at an elevation of 5100-5500~m above sea level, enjoying conditions comparable to the Chajnantor Plateau or better.}
    \label{fig:atmos}
\end{figure}

The exact AtLAST site will be determined after the testing activities to take place during the EU-funded Design study (see Section \ref{sec:WP3}), but will likely be in the Atacama Desert at an elevation between 5100-5500~m above sea level (a.s.l). Figure \ref{fig:atmos} shows that the median atmospheric transmission at the Chajnantor Plateau (5060~m a.s.l) is nearly comparable to that of Cerro Chajnantor (5600~m a.s.l), while offering a dramatic improvement at $\nu \gtrsim 400~\rm GHz$ over several other sites in use.
The excellent observing conditions at these altitudes necessitates an improved primary mirror alignment and surface accuracy of $\approx 20-25~\rm \mu m$ RMS in order to exploit atmospheric windows up to $\approx 350~\rm \mu m$.
While higher elevations correlate with drier, more transparent atmosphere, we note that construction and operations above 5500 meters would imply additional legal and logistic considerations \cite{Otarola2019}. %We therefore are largely exploring sites at elevations 5100-5500 meters above sea level.

\subsection{Proposed Instrumentation}

To develop requirements for each individual instrument that will enable us to achieve our entire wide range of science goals, we select a driving (i.e.\ the most challenging or stringent in terms of required instrument performance) operational goal for each instrument, and develop the technical requirements to meet that goal. We will mature this further during the design study, but at the moment we consider the following operational goals as driving: 1) full band spectral mapping of extended sources (more than a few arcminutes) as the driving goal for the high resolution spectrometer, 2) detection of high redshift galaxies for the continuum camera, and 3) redshift determination of detected galaxies for the low resolution spectrometer. Broadly, these three driving goals determine the frequency band allocation within the limited feed horn count of the high resolution spectrometer and limited focal plane area of the continuum camera, and also optimise the spectral resolution and frequency coverage for the low resolution spectrometer.

\begin{figure}
   \begin{center}
   \begin{tabular}{c} %% tabular useful for creating an array of images 
   \includegraphics[clip,trim=13mm 3mm 18mm 15mm, width=0.95\textwidth]{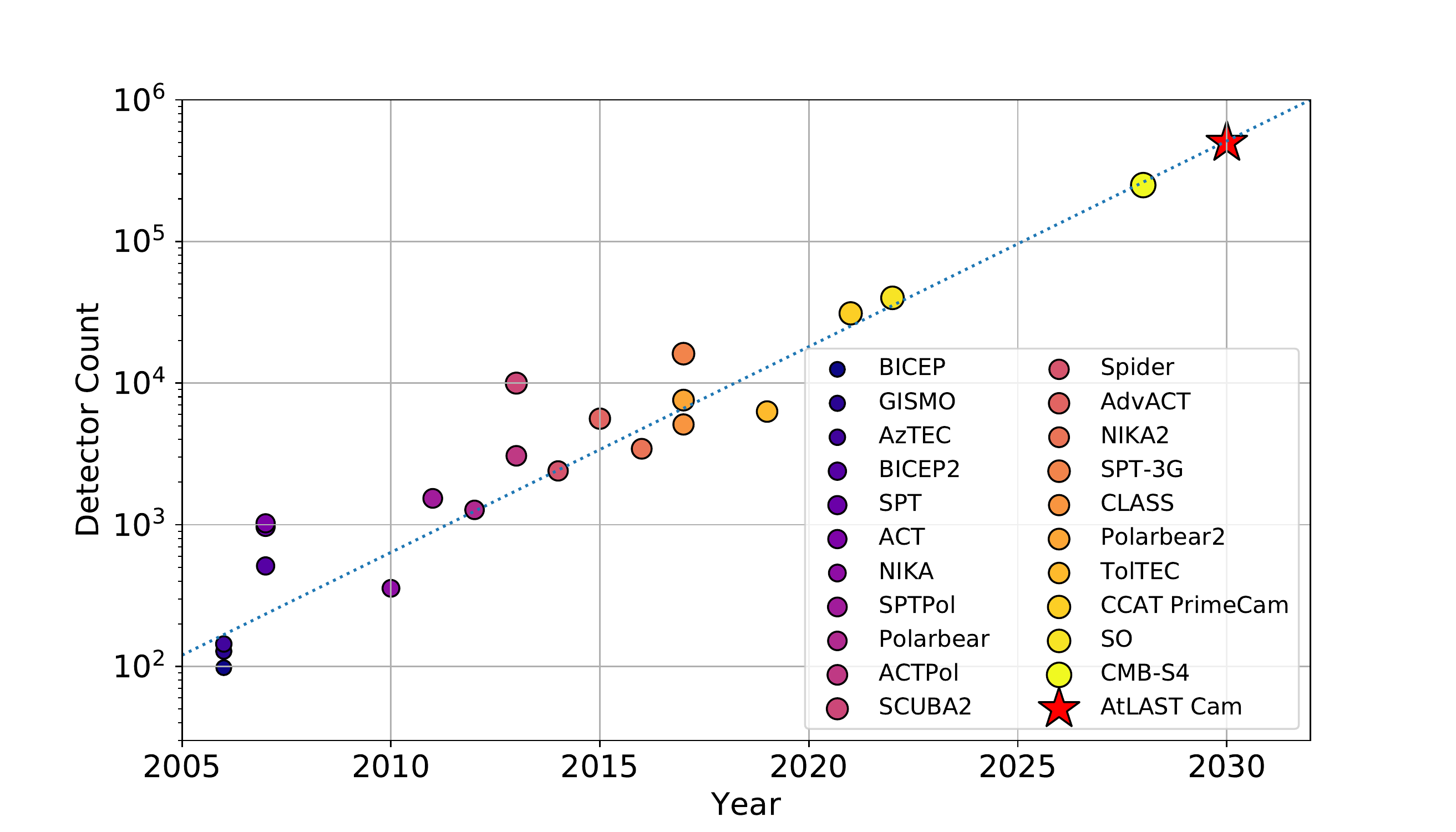}
   \end{tabular}
   \end{center}
   \caption[example] 
   { \label{fig:nbolos} 
Detector counts of several field leading millimetre or submillimetre-wave direct-detection instruments, using bolometers, transition edge sensors, or kinetic inductance detectors.  
The data were compiled from an amalgam of publications and websites, and are shown here solely to illustrate a trend.
The best-fit log-linear relation implies the number of detectors can increase by an order of magnitude roughly every seven years, reaching the megapixel regime circa 2032.   Our projection for a wide-field, Cassegrain-mounted first-generation camera, which we refer to generically as `AtLAST Cam,' is marked with a red star.  Acronyms are defined in Table \ref{tab:acronyms} in the appendix.}
\end{figure}

With these goals in mind, a few potential first-generation instruments are therefore:  
1) a heterodyne focal plane array\cite{Groppi2019} will provide high spectral ($\delta v < 1~\rm km~s^{-1}$) resolution, wide-field mapping of e.g.\ the Galactic Plane \cite{Klaassen2019,Stanke2019} and extra-galactic fields\cite{Geach2019}. 
2) a wide-field (order of a degree) multi-chroic continuum camera, referred to here as AtLAST Cam, likely exploiting kinetic inductance detector (KID) technologies \cite{Austermann2018,Bryan2018,Johnson2018}.  The science case for this is largely to provide photometry of extra-galactic fields, mapping of the primary and secondary effects of the CMB (e.g.\ the SZ effects and CMB lensing)\cite{Basu2019, Battistelli2019, Cicone2019, Dannerbauer2019, Mroczkowski2019, Mroczkowski2019b, Ruszkowski2019, Sehgal2019}, and would likely included polarimetry.  See Figure \ref{fig:nbolos} for how the detector counts required for AtLAST Cam compare to previous and planned direct-detection (sub)mm instruments.
3) a (sub)mm-wave integral field unit (IFU) relying on a direct detection spectrometer approach, delivering resolving power $R\approx300-1000$. While many small aperture instruments are pursuing an intensity mapping approach, the 50-m aperture of AtLAST will enable resolved tomography of the Universe out to the redshift of formation\cite{Battaglia2019, Geach2019}. Like the wide-field multi-chroic mapper/polarimeter above, this instrument will likely rely on KIDs. Example instrumental pathfinders are now abundant in the literature \cite{Shirokoff2012, Bryan2016, Barrentine2016, Cataldo2018, McGeehan2018, Endo2019, Endo2019b, CONCERTO, EXCLAIM2020, Karkare2020}.

\section{Next Steps: The EU-funded Design Study}
\label{sec:design_study}

\begin{comment}
    \item Scope
    \item Work Packages
    \item timescales
\end{comment}

The scope of the design study is to develop a fully-fledged, construction-ready telescope design based on the science requirements of the community. Because each funding agency has their own requirements for preliminary design review (PDR), the design documents stop just short of full PDR requirements.  This will allow us to incorporate specific requests as the project matures. The three year design study will official commence in March 2021.

\begin{figure}
    \centering
    \includegraphics[width=0.95\textwidth]{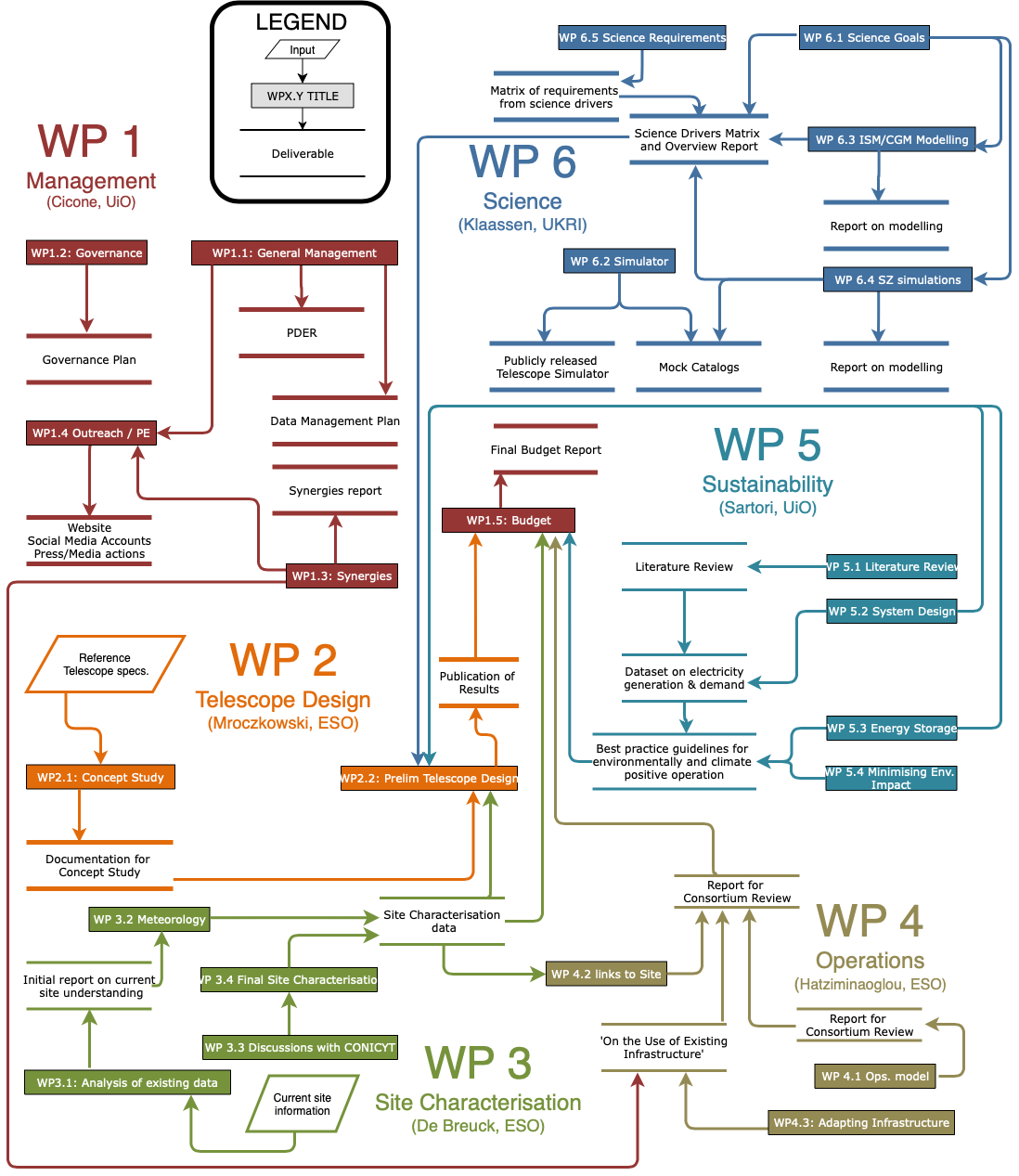}
    \caption{PERT chart showing the connections between the Horizon 2020 Design Study work packages and expected deliverables.}
    \label{fig:H2020PERT}
\end{figure}

The study is broken down into six work packages: Overall Management (WP1), Telescope Design (WP2), Site Selection (WP3), Operations (WP4), Sustainability (WP5) and Science Case Refinement (WP6). Each of these work packages is described in more detail below, with an overview presented in Figure \ref{fig:H2020PERT}.

\begin{comment}
    \item {\bf Overall Management (WP1)} coordinates the design study overall and aims to define a structure of governance for the project.
    \item {\bf Telescope Design (WP2)} aims to deliver a construction-ready, fully-engineered telescope design, with end-to-end simulations of the mechanical and systems interfaces.
    \item {\bf Site Selection (WP3)}
    \item {\bf Operations (WP4)}
    \item {\bf Sustainability (WP5)}
    \item {\bf Science Case Refinement (WP6)}
\end{comment}

\subsection{WP1: Overall Management}
\label{sec:WP1}
This work package, led by the University of Oslo, coordinates the design study overall and has the ultimate goal of delivering a plan for the long-term sustainability of AtLAST and its governance structure. The responsibilities of this package also include: supervising the activities of the different work packages, managing external collaborations and discussing future partnerships, investigating synergies with other astronomical facilities, and finalising the budget for the construction and operations of AtLAST. In addition, WP1 is in charge of communicating and disseminating the results of the project.

\subsection{WP2: Telescope Design}
\label{sec:WP2}
The second work package, led by researchers at ESO working with industrial partner OHB Digital Connect GmbH (formerly MT Mechatronics),\footnote{\url{www.ohb-digital.de}} is the largest individual package in the design study. The first task of this work package is to derive a baseline definition for the technical design of the telescope that will explore a wide range of design parameters and weigh their relative costs against their performance.%\footnote{\pk{AtLAST memo 1 link} \cc{CC: we need to produce such a document and link it to the new AtLAST website, currently we only have the (slightly redundant) meeting minutes document.}}. 
The remaining work will focus on delivering a construction-ready, fully-engineered conceptual telescope design report, including end-to-end simulations of the mechanical and systems interfaces.

\subsection{WP3: Site Selection}
\label{sec:WP3}
The third work package, led by researchers at ESO, aims at better characterising, and ultimately reporting on, suitable sites for the observatory. The first task of WP3 is analysing new and existing weather data for the Chajnantor Plateau area, by taking into account not only the parameters that determine the quality of sub-mm observations (e.g.\ precipitable water vapour, wind speed, and temperature) and obtaining new meteorological data such as the vertical wind profile.  It will additionally analyse the access to renewable energy resources and the environmental impact of the infrastructure. 
In parallel, WP3 will investigate the legal aspects of building a telescope within the Atacama Astronomy Park (Parque Astron\'omico de Atacama).

\subsection{WP4: Operations}
\label{sec:WP4}
WP4, led by researchers at ESO, will study scientific and technical operations models for the observatory, which will be distilled into an Operations Plan. All possible ramifications between on-site, completely remote observing and anything in between will be explored.  This work package will also address the scheduling and allocation of engineering/commissioning time.  As shown in Figure \ref{fig:H2020PERT}, WP4 is tightly linked with the site selection (WP3). The operations model will consider several factors such as the use of existing infrastructures, logistics and costs for transportation of staff and fuel, health and safety risks as well as the impact of the different operations schemes on gender, diversity and inclusiveness.
Following state-of-the-art business management practices, the Operations Plan will also include, among other things, a Continuity of Operations, a Crisis Communication, a Critical Infrastructure Protection, a Disaster Recovery and an Occupant Emergency Plan, for the eventuality of situations like the current pandemic. 
Finally, the responsibilities of WP4 include coordinating and collaborating with other observatories on Chajnantor (such as APEX and ALMA) for the use of common infrastructures during the construction of AtLAST, and, in the future, for the development and testing of its instrumentation.

\subsection{WP5: Sustainability}
\label{sec:WP5}
The Sustainability work package, led by the University of Oslo, is a first of its kind that is being considered by an observatory at the design study stage. WP5 will investigate the use of renewable-based energy systems for the telescope infrastructure, and explore their feasibility for providing energy to nearby communities that may not have a reliable electricity supply. One key component will be the study of a suitable hybrid energy storage system combining battery and hydrogen fuel cell systems to store energy derived from solar power plants. The energy management will be performed using both actual data as well as appropriate simulation models, in order to guarantee short- and long-term energy storage while maximising the performance of the systems components.

\subsection{WP6: Refinement of Science Case}
\label{sec:WP6}
Section \ref{sec:science}, as well as a large number of science white papers submitted to the U.S.\ Astro2020 Decadal Survey and Canadian Long Range Plan 2020, outline some of the key science drivers used to develop the case for this new observatory. 
Work Package 6, led by the UK Science and Technology Facilities Council (part of UK Research and Innovation), aims to distil a set of observatory requirements from the key science drivers - to refine the values presented in Table \ref{tab:telescope} based on how vital they are to meet the science goals.  It is the one that will draw the most from the expertise of the scientific community, particularly when working to mature the science case.
To best understand how changes in telescope design parameters affect the science cases, this work package also includes the creation of an observatory simulator / sensitivity calculator which will feature our working understanding of the light paths through the telescope, and toy models of the instruments outlined in Section \ref{sec:atlast} above, using sensitivities 
derived from similar existing instruments scaled up to the highly multiplexed versions expected for AtLAST. 
In addition, WP6 will deliver theoretical expectations and mock catalogues for AtLAST observations of the interstellar, circumgalactic and intergalactic media, in both local and distant galaxies.

\section{SUMMARY AND WAY FORWARD}
\label{sec:conclusions}

Here we have presented a project overview and the baseline design for the AtLAST 50-m telescope, a community-driven endeavour on which work will soon commence.  
The AtLAST design is driven by the need to provide: 1) sufficient resolution for astrophysical studies and for beating the confusion limit in its upper bands as well as sufficient overlap in the Fourier domain with the spatial scales probed by ALMA and the ACA, 2) and active primary (and possibly secondary) mirror,  with closed-loop metrology for focusing, and a surface quality optimised to take advantage of the excellent atmospheric transmission high in the Atacama Desert, 3) a transformative level of throughput, and 4) a superior spatial dynamic range possible with many millions of beams, fast scan speeds that modulate the atmosphere, and a far field response that ensures the atmospheric noise contribution is highly-correlated and can be removed as a common mode.
A salient feature of the AtLAST design is that it will deliver better resolution than the ACA, which has a maximum baseline of 45~m, while providing over $>4\times$ the collecting area ($\sim 1/3$ that of full ALMA) and a correspondingly high sensitivity, and a field of view thousands of times larger than either.  These key parameters will lead to an unprecedented leap in mapping speed capabilities that will define an entire generation of sub-mm astronomy.

We invite community participation in all the many facets of this large, transformative project to deliver the next-generation sub-mm observatory.

\acknowledgments % equivalent to \section*{ACKNOWLEDGMENTS}       
This project has received funding from the European Union’s Horizon 2020 research and innovation programme under grant agreement No 951815.

% References
\bibliography{report} % bibliography data in report.bib
\bibliographystyle{spiebib} % makes bibtex use spiebib.bst

%\clearpage
%\newpage
\appendix

\section{appendix}\label{sec:appendix}

\begin{center}
\begin{tabular}{|l|l|}
\hline
Acronym & Meaning \\
\hline
ACA	& Atacama Compact Array (Morita Array, 7-metre component of ALMA)\\
\href{https://act.princeton.edu/}{ACT}	& Atacama Cosmology Telescope\\
\href{https://act.princeton.edu/overview/camera-specifications/actpol}{ACTpol}	& Atacama Cosmology Telescope polarimetry receiver upgrade\\
\href{https://act.princeton.edu/overview/camera-specifications/advact}{AdvACT}& Advanced (3rd generation) receiver Atacama Cosmology Telescope\\
\href{https://www.ru.nl/astrophysics/black-hole/africa-millimetre-telescope/}{AMT} &  Africa Millimetre Telescope\\
\href{https://almaobservatory.org}{ALMA} &  Atacama Large Millimeter/submillimeter Array\\
\href{http://www.apex-telescope.org/}{APEX} & Atacama Pathfinder Experiment (APEX)\\
\href{https://atlast-telescope.org}{AtLAST} &  Atacama Large Aperture Submillimetre Telescope  \\
\href{https://www.nao.ac.jp/en/research/telescope/aste.html}{ASTE} & Atacama Submillimetre Telescope Experiment\\
\href{http://bicepkeck.org/}{BICEP} & Background Imaging of Cosmic Extragalactic Polarization\\
CCAT & Proposed 25-m telescope; name derived from ``Cerro Chajnantor Atacama Telescope''\\
\href{https://www.ccatobservatory.org/}{CCAT-p} & CCAT-prime, the 6-metre evolution of CCAT (see also FYST)\\
%CGM & circum-galactic medium\\
\href{https://sites.krieger.jhu.edu/class/}{CLASS} & Cosmology Large Angular Scale Surveyor\\
\href{https://cmb-s4.org/}{CMB-S4}	& The US Dept of Energy backed ``Stage IV'' proposed CMB experiment\\
\href{http://cso.caltech.edu/}{CSO} & Caltech Submm Observatory \\
\href{https://eventhorizontelescope.org/}{EHT} & Event Horizon Telescope\\
FBC	& Flexible Body Control\\
%FIR	& Far infrared\\
FoV	& Field of View\\
FYST & Fred Young Submm Telescope (formerly CCAT-prime)\\
\href{https://greenbankobservatory.org/science/telescopes/gbt/}{GBT}	& Green Bank Telescope\\
GISMO & Goddard IRAM Superconducting Millimeter Observer \\
%GLT & Greenland Telescope\\
\href{http://www.lmtgtm.org/}{GTM}	& Gran Telescopio Milimétrico (alternate name for the LMT)\\
IFU & integral field unit\\
\href{https://www.iram-institute.org}{IRAM} &	Institut de Radioastronomie Millimétrique\\
\href{https://www.eaobservatory.org/jcmt/}{JCMT} & James Clerk Maxwell Telescope\\
JWST &	James Webb Space Telescope\\
KID & Kinetic Inductance Detector\\
LASSI &	Laser Active Surface Scanning Instrument\\
\href{https://www.llamaobservatory.org/}{LLAMA} & Large Latin American Millimeter Array\\
\href{http://www.lmtgtm.org/}{LMT}	& Large Millimeter Telescope (alternate name for the GTM)\\
\href{https://www.lsst.org/}{LSST} & The Vera C. Rubin Observatory, formerly ``Large Synoptic Survey Telescope''\\
\href{https://www.lstobservatory.org/}{LST} & Large Submillimeter Telescope\\
MAO & Millimetric Adaptive (or Active) Optics\\
\href{https://modis.gsfc.nasa.gov/}{MODIS} & Moderate Resolution Imaging Spectroradiometer\\
\href{https://ngvla.nrao.edu/}{ngVLA} & The next-generation Very Large Array\\
NIKA & New IRAM KID Array\\
\href{https://www.iram-institute.org/EN/noema-project.php?ContentID=9&rub=9&srub=0&ssrub=0&sssrub=0}{NOEMA} & NOrthern Extended Millimeter Array\\
\href{https://origins.ipac.caltech.edu/}{OST}	& Origins Space Telescope\\
PDR & Preliminary Design Review\\
PI & Principal investigator \\
\href{https://www.cfa.harvard.edu/sma/}{SMA} & Submillimeter Array\\
\href{https://simonsobservatory.org/}{SO} & Simons Observatory\\
\href{https://en.wikipedia.org/wiki/SPICA_(spacecraft)}{SPICA} &	Space Infrared Telescope for Cosmology and Astrophysics\\
\href{https://pole.uchicago.edu/}{SPT}	& South Pole Telescope\\
SPT-3G	&  South Pole Telescope 3rd-Generation receiver\\
SPTpol	& South Pole Telescope polarimetry upgrade\\
SQUID & Superconducting Quantum Interference Detector\\
\href{http://www.ioa.s.u-tokyo.ac.jp/TAO/en/}{TAO} & (University of) Tokyo Atacama Observatory \\
TES & Transition Edge Sensor\\
%TLR & Top Level Requirement\\
\hline
\end{tabular}
\label{tab:acronyms}
\end{center}

\end{document}